\documentclass[twocolumn,showpacs,preprintnumbers,amsmath,amssymb,prb,superscriptaddress]{revtex4}

\usepackage{graphicx}
\usepackage{float}

\begin{document}

\title{P-wave Cooper pair splitting}
\author{H. Soller$^1$ and A. Komnik}
\affiliation{Institut f\"ur Theoretische Physik,
Ruprecht-Karls-Universit\"at Heidelberg,\\
 Philosophenweg 19, D-69120 Heidelberg, Germany}
\date{\today}

\begin{abstract}
Splitting of Cooper pairs has recently been realized experimentally for s-wave Cooper pairs. A split Cooper pair represents an entangled two-electron pair state which has possible application in on-chip quantum computation. Likewise the spin-activity of interfaces in nanoscale tunnel junctions has been investigated theoretically and experimentally in recent years. However, the possible implications of spin-active interfaces in Cooper pair splitters so far have not been investigated.\\
We analyse the current and the cross correlation of currents in a superconductor ferromagnet beamsplitter including spin-active scattering. Using the Hamiltonian formalism we calculate the cumulant generating function of charge transfer. As a first step, we discuss characteristics of the conductance for crossed Andreev reflection in superconductor ferromagnet beamsplitters with s-wave and p-wave superconductors and no spin-active scattering. In a second step, we consider spin-active scattering and show how to realize p-wave splitting only using a s-wave superconductor via the process of spin-flipped crossed Andreev reflection. We present results for the conductance and cross correlations.\\
Spin-activity of interfaces in Cooper pair splitters allows for new features in ordinary s-wave Cooper pair splitters, that can otherwise only be realised by using p-wave superconductors. In particular it provides access to Bell states different from the typical spin singlet state.
\end{abstract}


\maketitle

\section{Introduction}
Solid state entanglers represent the electronic analogues to the optical setups used for Bell inequality tests. Most setups propose a superconductor as a source of spin-entangled s-wave Cooper pairs \cite{PhysRevB.63.165314,PhysRevB.65.165327,PhysRevLett.88.197001}. These are transferred to spatially separated leads by the process of Crossed Andreev Reflection (CAR) \cite{PhysRevLett.74.306,deutscher:487}.
Different setups have been considered in order to achieve CAR without being dominated by elastic cotunnelling or ordinary Andreev Reflection (AR) processes \cite{springerlink:10.1007/s100510051010,PhysRevB.77.094512,PhysRevB.66.161320,sidorenko}. S-wave Cooper pairs are entangled in energy and spin space. Therefore, one may either filter the electrons of a Cooper pair in spin space (using ferromagnets \cite{0295-5075-81-4-40002,PhysRevLett.93.197003,springerlink:10.1007/s00339-007-4193-4} or Luttinger liquids \cite{PhysRevB.65.165327}), or in energy space (using quantum dots \cite{PhysRevB.63.165314,PhysRevB.83.125421}, coupling to an electromagnetic mode \cite{2007NatPh...3..455Y}, an appropriate voltage bias \cite{0295-5075-67-1-110,springerlink:10.1007/s10051-001-8675-4,PhysRevLett.95.027002,PhysRevB.79.054505} or ac-bias \cite{0295-5075-86-3-37009}). Filtering using quantum dots with large onsite interaction has been successfully realized in experiment \cite{PhysRevLett.104.026801,19829377,2012arXiv1204.5777S} and a non-local resistance has been measured as characteristics of Cooper pair splitting. Moreover, a positive cross correlation measured in a superconductor and normal-metal three-terminal device gave compelling evidence for CAR \cite{2009arXiv0910.5558W}. However, s-wave Cooper pairs only give access to one of the Bell states, namely $|\psi^{-}\rangle = 1/\sqrt{2} \; (|\uparrow\rangle_1 \otimes |\downarrow\rangle_2 - |\downarrow\rangle_1 \otimes |\uparrow\rangle_2)$, where the subscripts 1 and 2 refer to two normal leads. P-wave Cooper pairs give access to the other Bell states, especially those like $|\phi^{\pm}\rangle = 1/\sqrt{2} \; (|\downarrow\rangle_1 \otimes |\downarrow\rangle_2 \pm |\uparrow\rangle_1 \otimes |\uparrow\rangle_2)$, involving fully spin-polarized combinations of the two electrons. Thus, a p-wave Cooper pair splitter represents the essential counterpart to s-wave Cooper pair splitters as on-chip sources of spin-entangled Einstein-Podolsky-Rosen (EPR) electron pairs \cite{PhysRev.47.777,RevModPhys.81.1727,2011arXiv1105.2583H}.\\
In this paper we approach the problem of p-wave Cooper pair splitting in two steps. First, we show that p-wave splitting may easily be identified in a hybrid junction between a superconductor and two ferromagmets (setup shown in Fig. \ref{sfffull}). However, p-wave superconductors that can be easily handled in quantum transport experiments are presently not available. Therefore, in a second step we show how p-wave splitting can be realized without using a p-wave superconductor. Indeed, the previous works on the charge transfer statistics of superconductor ferromagnet beamsplitters \cite{soller,0295-5075-81-4-40002,PhysRevB.78.224515} neglected the recently predicted \cite{PhysRevB.81.094508,PhysRevLett.101.257001,PhysRevB.77.064517} and observed \cite{2010arXiv1012.3867H,PhysRevB.83.081305} effects of the interface on charge transfer. By including a better description of the interface one can describe the shared triplet pairs formation between the ferromagnets. We calculate the Full Counting Statistics (FCS) which allows us to identify the corresponding charge transfer process and to calculate the cross correlation as an experimentally observable quantity \cite{2009arXiv0910.5558W}.\\

\section{Results and Discussion}
\subsection{Superconductor Ferromagnet Beamsplitters}
Splitting of spin-polarized p-wave Cooper pairs can easily be identified in the conductance. From the result in \cite{soller} we find the generalization of Beenakker's formula \cite{beenakker} for zero-bias conductance of a beamsplitter realized by a resonant level between a superconductor and two ferromagnets
\begin{eqnarray}
G_{CAR}(\sigma_1, \sigma_2) &=& \frac{4e^2}{h} \left[4 \Gamma^2 \Gamma_S^2 (1+ \sigma_1 P_1) ( 1+ \sigma_2 P_2)\right] \nonumber\\
&& \times \left[\Delta_d^2 + \Gamma_S^2 +  \Gamma^2 (2+ \sigma_1 P_1 + \sigma_2 P_2)^2\right]^{-1} \nonumber\\
&& \times \left[\Delta_d^2 + \Gamma_S^2 + \Gamma^2 (2- \sigma_1 P_1 - \sigma_2 P_2)^2\right]^{-1}, \label{gcar}
\end{eqnarray}
\begin{figure*}
\centering
\caption{Summary of the possible charge transfer processes in a superconductor ferromagnet beamsplitter. The superconductor (blue) and the two ferromagnets (red, are assumed to be fully polarized) are coupled via a quantum dot realised by an InAs nanowire (brown). The polarisation is indicated by green arrows. In the upper part, the situation for s-wave superconductors is shown. The Cooper pair may split if the two ferromagnets are antiparallely polarized. In the lower part the reversed situation for spin-polarized p-wave superconductors is depicted. The Cooper pair may now split only if the two ferromagnets are equally polarized.}
\label{sfffull}
\includegraphics[width=13cm]{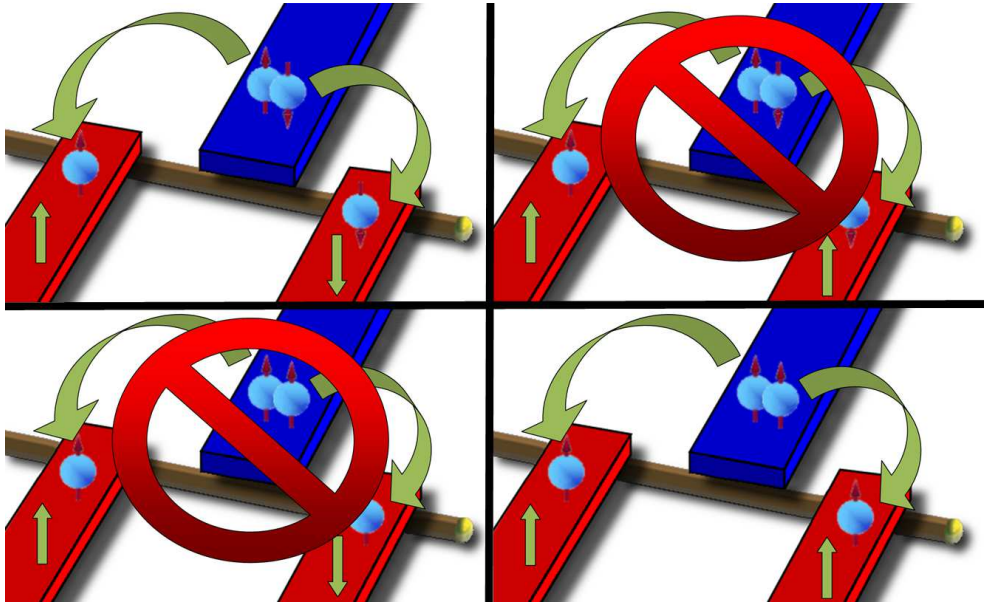}
\end{figure*}
where $\Gamma$ is the tunnel rate through the two barriers between the quantum dot and the ferromagnet (both are assumed to be equal) and $\Gamma_S$ is the tunnel rate between the superconductor and the quantum dot. $\Delta_d$ refers to the energy of the resonant level. $P_1$ and $P_2$ are the (parallel) polarisations of the ferromagnets and $\sigma_1$ and $\sigma_2$ are the spins of the electrons in a Cooper pair.\\
In usual s-wave superconductivity the spin directions obey $\sigma_1 = - \sigma_2$, and we may maximize CAR by choosing $P_1 = 1 = - P_2$ (or vice versa) \cite{0295-5075-81-4-40002}. However, if we choose the polarisations of the ferromagnets to be equal (i.e. $P_1 = 1 = P_2$) $G_{CAR}$ in Equation (\ref{gcar}) becomes zero (see Fig. \ref{sfffull}). The situation is reversed if we introduce a spin-polarized p-wave superconductor so that $\sigma_1 = \sigma_2$. Now splitting is maximized if $P_1 = P_2$. Of course now for antiparallel polarisation the current is blocked (see 
Fig. \ref{sfffull}). Therefore p-wave splitting is easily identified in the crossed conductance. The assumption of maximal polarisation is not easy to realize in experiment \cite{PhysRevLett.105.207001} but we use it to illustrate the argument.\\
In spite of the apparent simplicity of this consideration a serious problem remains: p-wave superconductors are very rare and materials such as $\mbox{Sr}_2\mbox{Ru}\mbox{O}_4$ are hard to handle \cite{PhysRevLett.93.167004}.\\
However, recent theoretical and experimental progress showed that the treatment of superconductor - ferromagnet interfaces requires special care with respect to the exact form of the interface and the associated spin-active nature of tunnelling \cite{PhysRevB.80.184511}. In the rest of this work we want to show how a spin-active interface can be used in exactly the same way as a p-wave superconductor and therefore allows to generate the other Bell states. In order to accomplish this goal we will show how the transport characteristics of p-wave splitting can be imitated by spin-active scattering.

\subsection{Cumulant generating function with spin-active scattering}
To identify the separate charge transfer processes and to evaluate the conductances and cross correlations we calculate the FCS of charge transfer \cite{nazarov,levitov-1996-37} using the generalized Keldysh technique \cite{PhysRevB.73.195301,PhysRevLett.103.136601,PhysRevB.80.041309}. The aim is to calculate the Cumulant Generating Function (CGF) $\chi(\lambda)$ of the probability distribution $P(q)$ to transfer $q$ units of charge during a (long) measurement time $\tau$: $\chi(\lambda) = \sum_q e^{i \lambda q} P(q)$. Partial derivatives of this function provide direct access to the cumulants (irreducible moments) of $P(q)$.\\
We model the superconductor ferromagnet beamsplitter by two ferromagnets $F1$ and $F2$ which are tunnel coupled to a resonant level, which is the simplest model of a quantum dot. The superconductor is also coupled to the resonant level via $H_{T1}$ and $H_{T2}$ which takes into account the interface effects. The result is the Hamiltonian
\begin{eqnarray}
H &=& H_{F1} + H_{F2} + H_{TF1} + H_{TF2} + H_S \\
&& + H_{T1} + H_{T2} + H_d, \; \mbox{where} \label{hsys}\\
H_d &=& \sum_\sigma \Delta_d \tilde{d}_\sigma^+ \tilde{d}_\sigma,
\end{eqnarray}
represents the resonant level. Throughout the rest of exposition we use units in which $e = \hbar = k_B = 1$ and restrict ourselves to a quantum dot on resonance $\Delta_d = 0$.\\
Ferromagnetic electrodes are described in the language of electron field operators $\Psi_{k,\alpha,\sigma}$, where $\alpha = F1, F2$ using the Stoner model with an exchange energy $h_{ex}$ as in \cite{melin}
\begin{eqnarray}
&& H_\alpha = \sum_{k, \sigma} \epsilon_k \Psi_{k,\alpha,\sigma}^+ \Psi_{k,\alpha, \sigma}\nonumber\\
&& - h_{ex, \alpha} \sum_k (\Psi_{k,\alpha,\uparrow}^+ \Psi_{k,\alpha,\uparrow} - \Psi_{k,\alpha,\downarrow}^+ \Psi_{k,\alpha,\downarrow}). \label{ferro}
\end{eqnarray}
Consequently they can be described as fermionic continua with a spin-dependent DOS $\rho_{0,\alpha,\sigma} = \rho_{0,\alpha} (1+ \sigma P_{\alpha})$, where $P_{\alpha}$ is the polarisation. The superconductor is described using the ordinary s-wave BCS Hamiltonian
\begin{eqnarray}
H_S &=& \sum_{k,\sigma} \epsilon_k \Psi_{kS\sigma}^+ \Psi_{kS\sigma} \nonumber\\
&& + \Delta \sum_k (\Psi_{kS\uparrow}^+ \Psi_{-kS\downarrow}^+ + \Psi_{-kS\downarrow} \Psi_{kS\uparrow}). \label{super}
\end{eqnarray}
We express the energies of the dot and the reservoirs relative to the superconductor chemical potential \cite{PhysRevB.50.3982} so that $\mu_S = 0$ and $V_{\alpha} = \mu_S - \mu_{\alpha} = - \mu_{\alpha}$ is the chemical potential of the ferromagnets. The quasiparticle density of states in the superconductor has the form: $\rho_S = \rho_{0S} |\omega|/\sqrt{\omega^2 - \Delta^2}$, where $\rho_{0S}$ is constant in the wide-band limit. Tunnelling between the superconductor and the quantum dot is given by \cite{PhysRevLett.8.316}
\begin{eqnarray}
H_{TS} = \sum_\sigma \tilde{\gamma}_{S} [\tilde{d}_\sigma^+ \Psi_{S,\sigma} (x=0) + \mbox{H.c.}], \label{ts}
\end{eqnarray}
where $\tilde{\gamma}_{S}$ is the corresponding tunnelling amplitude between the dot and the superconductor. The tunnelling is assumed to be local and to occur at $x=0$ in the coordinate system of the respective electrode. $\Psi_{S,\sigma} (x)$ refers to the electron field operator of the superconductor introduced in Equation (\ref{super}) in position space.\\
Finally, we need a Hamiltonian approach \cite{PhysRevB.54.7366} for spin-active scattering. There are manifold effects like spin-orbit coupling, magnetic anisotropy or spin relaxation that give rise to spin-activity of interfaces \cite{PhysRevB.81.094508}. Previous studies of point contacts used a scattering states description \cite{PhysRevB.81.094508,PhysRevLett.101.257001,PhysRevB.77.064517,PhysRevLett.102.227005} in order to introduce a spin-active scattering angle as a phenomenological parameter to characterise the interface or a wave-function matching technique \cite{PhysRevB.81.174526,PhysRevB.83.054513}. We adopt the simplest possible approach and follow \cite{PhysRevLett.89.286803,PhysRevLett.90.116602,yamada} by introducing two tunnelling Hamiltonians at each of the interfaces between the quantum dot and a ferromagnet
\begin{eqnarray}
H_{TF1,\alpha} &=& \sum_\sigma \tilde{\gamma}_{\alpha,1} [\tilde{d}_\sigma^+ \Psi_{\alpha,\sigma}(x=0) + \mbox{H.c.}],\nonumber\\
H_{TF2,\alpha} &=& \sum_\sigma \tilde{\gamma}_{\alpha,2} [\tilde{d}_\sigma^+ \Psi_{\alpha,-\sigma}(x=0) + \mbox{H.c.}]. \label{tf}
\end{eqnarray}
$H_{TF1,\alpha}$ describes normal spin-conserving tunnelling whereas $H_{TF2,\alpha}$ refers to the spin-flip processes [$\Psi_{\alpha,-\sigma}(x=0)$ represents an electron in the ferromagnet with opposite spin]. If we take spin-active scattering into account this way we have 5 different tunnel couplings. In order to reduce the number of parameters and for clarification of the discussion of spin-active scattering effects we want to limit ourselves to a special constellation of parameters as far as spin-active scattering is concerned, namely $\gamma_{F1,1} = \gamma_{F2,1}, \; \gamma_{F1,2} = \gamma_{F2,2}$. In this case we define operators
\begin{eqnarray}
d_\sigma = \frac{\gamma_{F1,1} d_\sigma + \gamma_{F1,2} d_{-\sigma}}{\sqrt{\gamma_{F1,1}^2+ \gamma_{F1,2}^2}}, \label{dop}
\end{eqnarray}
which allow us to rewrite the tunnelling Hamiltonians in Equations (\ref{ts}) and (\ref{tf}) as
\begin{eqnarray}
H_{TF,\alpha} &=& \sum_\sigma \gamma_{\alpha} [d_\sigma^+ \Psi_{\alpha,\sigma} (x=0) + \mbox{H.c.}], \label{t1}\\
H_{T1} &=& \sum_\sigma \gamma_{S,1} [d_\sigma^+ \Psi_{S,\sigma}(x=0) + \mbox{H.c.}],\nonumber\\
H_{T2} &=& \sum_\sigma \gamma_{S,2} [d_\sigma^+ \Psi_{S,-\sigma}(x=0) + \mbox{H.c.}]. \label{t2}
\end{eqnarray}
This minimal model still reveals all the transport properties we want to discuss here.\\
In order to access the CGF, we introduce the counting fields for the leads attached to the quantum dot as time-dependent fictitious parameters aka counting fields in front of the creation and annihilation operators of the respective electrodes
\begin{eqnarray}
\Psi_{S,\sigma}(x=0) &\rightarrow& e^{-i \lambda_S(t)/2} \Psi_{S,\sigma}(x=0),\nonumber\\
\Psi_{\alpha,\sigma} (x=0) &\rightarrow& e^{-i \lambda_{\alpha}(t)/2} \Psi_{\alpha,\sigma}(x=0). \label{sub}
\end{eqnarray}
$\lambda_{(S,1,2)}(t)$ takes the value $+(-) \lambda_{(S,1,2)}$ on the forward/backward branch of the Keldysh contour ${\cal C}$. The counting fields are only switched on during the measurement time $\tau$. In the limit of long measurement times $\tau$ the CGF can be calculated analytically using a generalized Green's function formalism that has previously been used to calculate the FCS of other quantum impurity systems \cite{PhysRevB.83.085401,PhysRevB.82.121414,PhysRevB.82.165441}. Following \cite{PhysRevB.70.115305} the CGF is given by the expectation value
\begin{eqnarray}
\ln \chi_{\mathrm{SFF}}(\vec{\lambda}, \tau) &=& \left \langle T_{\cal C} \exp \left[-i \int_{\cal C} \left( T_{TF1}^{\lambda_1} + T_{TF2}^{\lambda_2} \right. \right. \right. \nonumber\\
&& \left. \left. \left. + T_{T1}^{\lambda_S} + T_{T2}^{\lambda_S}\right)\right]\right\rangle, \label{cgf}
\end{eqnarray}
where $\vec{\lambda} = (\lambda_1, \lambda_2, \lambda_S)$ and $T_{\cal C}$ denotes the contour time ordering operator. $T_{TF1}^{\lambda_1}, \cdots, T_{T2}^{\lambda_S}$ are abbreviations for the tunnelling operators introduced in Equations (\ref{t1}) and (\ref{t2}) in combination with the substitutions defined in Equation (\ref{sub}).\\
For the FCS calculation in the limit of large measurement times (as assumed here) we follow \cite{PhysRevB.73.195301} and write the above expression using an adiabatic potential \cite{PhysRevLett.26.1030} which, in the limit of infinitely long measuring time, does not depend on it
\begin{eqnarray*}
\ln \chi_{\mathrm{SFF}}(\vec{\lambda}) = -i \int_0^\tau dt U (\vec{\lambda}) = - i \tau U(\vec{\lambda}).
\end{eqnarray*}
In turn the adiabatic potential is related to the counting field derivatives of the $T^\lambda$'s introduced in Equation (\ref{cgf})
\begin{eqnarray*}
\sum_{j = S,1,2}  \frac{\partial U}{\partial \lambda_{j = S,1,2}} &=& \left \langle \frac{T_{TF1}^{\lambda_1}}{\partial \lambda_1} \right \rangle_\lambda + \left \langle \frac{T_{TF2}^{\lambda_2}}{\partial \lambda_2} \right \rangle_\lambda \\
&& + \left \langle \frac{T_{T1}^{\lambda_S}}{\partial \lambda_S} \right \rangle_\lambda + \left \langle \frac{T_{T2}^{\lambda_S}}{\partial \lambda_S} \right \rangle_\lambda,
\end{eqnarray*}
where $\langle \cdot \rangle_\lambda$ is defined as $\langle \cdot \rangle_\lambda := 1 / \chi_{\mathrm{SFF}}(\vec{\lambda},\tau) \langle \cdot \rangle$ with $\langle \cdot \rangle$ being the ordinary expectation value with respect to the systems Hamiltonian in Equation (\ref{hsys}) with the tunneling Hamiltonians rephrased using the substitution defined in Equation (\ref{sub}). The derivatives of the $T^\lambda$'s take the form of mixed and $\vec{\lambda}$-dependent Green's function, e.g.
\begin{eqnarray*}
\left \langle \frac{T_{TF1}^{\lambda_1}}{\partial \lambda_1} \right \rangle_\lambda = \frac{-i \gamma_1}{2} \sum_\sigma  \left[e^{-i\lambda_1/2} \langle T_{\cal C} d_\sigma^+ \Psi_{1,\sigma}\rangle_\lambda + \mathrm {h.c.}\right].
\end{eqnarray*}
These mixed Green's functions have to account for all orders of the tunneling coupling. By expanding them to first order in the relevant tunnel coupling they can all be rewritten as a product of a bare lead Green's function (subscript $0$) and an exact-in-tunneling dot Green's function. E.g. for the derivative considered above
\begin{eqnarray*}
\langle T_{\cal C} d_\sigma^+ \Psi_{1,\sigma}\rangle_\lambda \propto \gamma_{1} \langle \Psi_{1,\sigma} \Psi_{1,\sigma}^+\rangle_0 \langle d_\sigma d_\sigma^+ \rangle_\lambda .
\end{eqnarray*}
Therefore the remaining task is an exact calculation of the dot Green's function. Since the system's Hamiltonian in Equation (\ref{hsys}) is quadratic the Dyson equation can be solved exactly. However, as also noted in \cite{PhysRevB.76.184510}, coupling of the dot to the superconductor automatically leads to the appearance of anomalous dot Green's functions of the type $\langle d_\uparrow^+ d_\downarrow^+ \rangle_\lambda$ to second order in the tunnel coupling to the superconductor. Additionally, the spin-flipping tunnel contribution in Equation (\ref{t2}) gives rise to correlation functions of the type $\langle d_\uparrow^+ d_\uparrow^+ \rangle_\lambda$ \cite{PhysRevLett.102.107008}. Consequently the full dot Green's function for the spin species $\uparrow$ becomes a 4-component vector of correlation functions $(\langle d_\uparrow d_\uparrow^+ \rangle_\lambda, \langle d_\uparrow d_\downarrow^+ \rangle_\lambda, \langle d_\uparrow^+ d_\downarrow^+\rangle_\lambda, \langle d_\uparrow^+ d_\uparrow^+ \rangle_\lambda)$. Since the counting fields take different signs on the backward/forward branch of the Keldysh contour each correlation function is a 2x2 Keldysh matrix again. The bare dot and lead Green's functions are e.g. given in \cite{PhysRevB.80.184510}. Since the solution of the Dyson equation allows to sum up all orders in the tunnelling couplings the result is exact and also valid at finite temperature as discussed also in \cite{PhysRevB.54.7366}.\\
However, the result for the average in Equation (\ref{cgf}) is quite long. We want to restrict ourselves to the study of the relevant aspects of the CGF in view of the possbility of p-wave splitting only.\\
Comparing our method to previous treatments of superconductor hybrids we should emphasize that also in the quasiclassical Green's function formalism \cite{1999SuMi...25.1251B} calculations of the current exact in tunneling have been done \cite{PhysRevB.75.172503,PhysRevB.76.224506}. However, there it was assumed to have two tunnel contacts to normal leads with large spatial separation. In our case the three leads are coupled via a quantum dot which corresponds to the opposite limit and involves an energy-dependent transmission. Therefore the setup considered here is closer related to the studies in Refs. \cite{PhysRevB.76.184510} and \cite{PhysRevB.74.214510} where, however, a disordered quantum dot and no spin-active scattering was assumed, whereas here we treat the ballistic case with spin-active scattering. Disorder for the case of tunnel contacts was considered using the Keldysh-Usadel formalism in Refs. \cite{PhysRevB.74.214512}, \cite{PhysRevLett.103.067006} and \cite{PhysRevB.82.024522}. Additionally we want to point out that the description of spin-active scattering used here and in Refs. \cite{PhysRevB.76.224506} and \cite{PhysRevB.81.094508} is different: in the latter works the spin-mixing angle is introduced as a phenomenological parameter whereas we use a second tunneling transparency to account for spin flips. For a comparison of both descriptions of tunnel contacts we refer the reader to \cite{soller3}.

\subsection{Spin-flipped Crossed Andreev Reflection}
\begin{figure*}
\centering
\includegraphics[width=17cm]{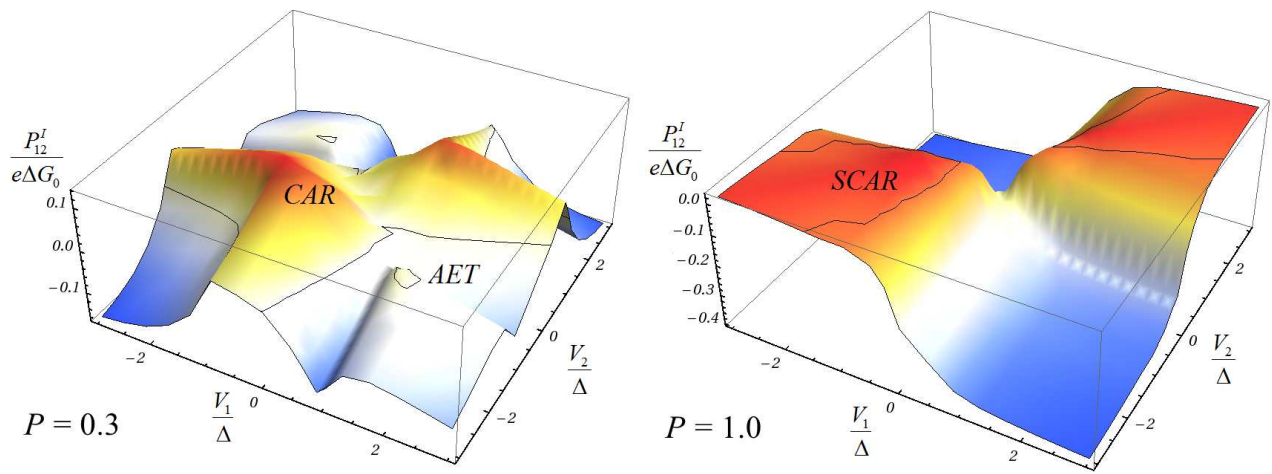}
\caption{Cross correlation of currents for different parameters: The cross correlation $P_{12}^I$ according to Equation (\ref{cross}) is calculated for two different sets of parameters. The polarisation of the ferromagnets is assumed to be equal in both cases. The left graph shows the result for $P=0.3$, $\Gamma_{S} = 4\Delta$, $\Gamma_{SF} = 0$, $\Gamma_{F1}= 0.4 \Delta$, $\Gamma_{F2} = 0.1 \Delta$ and $T=0.1\Delta$. The right graph is for $P=1$, $\Gamma_S = 2 \Delta$, $\Gamma_{SF} = \Delta$, $\Gamma_{F1} = 0.05 \Delta = \Gamma_{F2}$ and $T=0.1 \Delta$. A distortion of the superconducting DOS described by a Dynes parameter \cite{PhysRevLett.41.1509} of $\Gamma_D = 0.005 \Delta$ has been introduced in order to circumvent numerical artifacts of the diverging superconductor DOS. The cross correlations are positive in the deep red regions.}
\label{fig:combined}
\end{figure*}

In simple superconductor-ferromagnet tunnelling junctions the presence of spin-active scattering gives rise to a new type of AR. In ordinary AR an electron is retroreflected as a hole with opposite spin since Cooper pairs represent spin singlets. However, due to the spin-active nature of tunnelling in the setup considered here the hole or the electron spin can be flipped. This Spin-flipped Andreev Reflection (SAR) induces triplet correlations in the ferromagnet \cite{PhysRevB.81.094508}. To see whether similar effects occur in our setup we calculate the zero-bias conductance analogously to Equation (\ref{gcar}) for the case of $P_1 = P_2 = 1$, meaning maximal parallel polarisation
\begin{eqnarray*}
G_{SCAR} = \frac{16e^2}{h} \frac{4\Gamma^2 \Gamma_{SF}^2(4 \Gamma_{SF}^2 - \Gamma_S^2)^2}{[\Gamma_S^4 - 8 \Gamma_S^2 \Gamma_{SF}^2 + 16 \Gamma_{SF}^2(4 \Gamma^2 + \Gamma_{SF}^2)]^2}.
\end{eqnarray*}
$\Gamma_S = \pi \rho_{0s} (\gamma_{s1}^2+ \gamma_{s2}^2)$ and $\Gamma_{F,\alpha} = \pi \rho_0 \gamma_{\alpha}^2 =: \Gamma$ again refer to the tunnel rates for ordinary single electron tunnelling. $\Gamma_{SF} = \pi \rho_{0s} \gamma_{s1} \gamma_{s2}$ describes the additional spin-flip tunnelling rate at the interface. Obviously, spin-active scattering has lifted the current blocking indicated in Fig. \ref{sfffull} for a s-wave superconductor connected to two maximally parallelly polarised ferromagnets. Therefore this finite conductance is similar to the one obatined for a p-wave superconductor junction in Equation (\ref{gcar}). This indicates that this conductance for voltages below the gap is indeed associated to a Spin-flipped Crossed Andreev Reflection (SCAR) in which a triplet pair is transferred to the ferromagnets.\\
Of course one may also obtain this information from the CGF itself. However, the expression is complicated and the probability distribution of charge transfer is also not easy to access in an experiment \cite{epub12308,reulet-2003-91}. Therefore we follow \cite{PhysRevLett.88.197001} and use the cross correlation as an indication to probe whether the two charges of a Cooper pair transferred in a SCAR process are indeed transferred to the ferromagnets at the same time. The presence of non-zero $G_{SCAR}$ at voltages below the gap and $T=0$ in combination with a positive cross correlation can only be explained by a simultaneous transfer of a triplet pair to the ferromagnets, which implies that we indeed observe p-wave splitting. The cross correlation can be calculated as a mixed second derivative of the CGF
\begin{eqnarray}
P_{12}^I = - \left.\frac{1}{\tau} \frac{\partial^2 \ln \chi_{SFF}(\vec{\lambda} , \tau)}{\partial \lambda_1 \partial \lambda_2}\right|_{\vec{\lambda} \rightarrow 0}. \label{cross}
\end{eqnarray}
In Fig. \ref{fig:combined} the result of Equation (\ref{cross}) is shown for two different configurations of the couplings and polarisations. For moderate polarisation ($P=0.3$) and no spin-active scattering we find two cases in which positive cross correlation can be observed in accordance with previous results \cite{PhysRevB.82.014510,soller}. First, for voltages close to the superconducting gap and $V_1 \approx - V_2$ CAR is strongly suppressed and one expects single electron transmission to be dominant. Nonetheless, the energy-dependent DOS of the superconductor leads to large transmission coefficients for double AR from one ferromagnet to the superconductor and further to the second ferromagnet. This process is known as Andreev reflection Enhanced Transmission (AET) \cite{soller}. The second case of positive cross correlation can be observed for one bias voltage being close to zero and finite bias on the second electrode. In this case, CAR dominates over single electron transmission and induces positive cross correlation \cite{soller}.\\
This picture changes dramatically if we go over to the case of full polarisation ($P=1$) and finite spin-active scattering. Since AET relies on double AR, and consequently the spin symmetry of AR, it must dissapear since SAR violates spin-symmetry and it would be the only possible charge transfer process for a single lead for $P=1$. However, for $V_1 \approx V_2$ a positive cross correlation remains. This is exactly the position where we assume SCAR to be dominant since $V_1 \approx V_2$ means that single electron transfer between the ferromagnets does not occur.\\
The effect is, of course, still observable for $P<1$ but the polarisation should be rather strong. Spin-active scattering in superconductor-ferromagnet hybrids is a general phenomenon \cite{2010arXiv1012.3867H} and full polarisation was just assumed for clarity. Therefore, we believe that SCAR is a generic phenomenon that should also be present in superconductor ferromagnet beamsplitters with tunnel contacts \cite{springerlink:10.1007/s100510051010,PhysRevB.76.224506} or chaotic cavities \cite{PhysRevB.74.214510,PhysRevB.76.184510} since its origin does not lie in the precise form of the energy dependence of the transmission coefficients.\\

Experiments in the direction of the above described proposal have already been realized \cite{PhysRevB.81.024515}. Multi-terminal hybrid systems with embedded quantum dots \cite{PhysRevLett.95.126603} also based on InAs nanowires \cite{2011arXiv1105.2583H,PhysRevLett.104.246804} have already been realised experimentally. In such devices a new subgap structure has been observed that can be explained by SAR \cite{soller3}. This is of special importance, since in our consideration we did not include effects of Coulomb interaction on the dot and so we should worry about a possible a suppression of SCAR. The mean field analysis of \cite{soller3}, however, revealed that SAR and therefore also SCAR should be observable also in presence of strong Coulomb interaction. Apart from that, we can argue that also in interacting systems characteristic resonances, as the one of the resonant level considered here, are present and have a characteristic location and width associated with interactions. Therefore the general scenario should be robust. However, one should mind that for more extended quantum dots or nanowires disorder could play an important role as mentioned in \cite{PhysRevB.74.214512} and \cite{PhysRevLett.103.067006}. Another experiment has realized a superconductor-ferromagnet-superconductor junction based on Al and Co electrodes with two closely spaced cobalt wires bridging two aluminum electrodes \cite{2011arXiv1111.5415C}. In this experiment the resistancee in the case of antiparallel and parallel magnetisation of the two wires has been measured and for low temperatures it has been found that the antiparallel arrangement may even have higher resistance than the parallel one giving reliable evidence for spin-active scattering being present in the device. Concerning a possible experimental realisation using quantum dots one should mind that interaction should be small enough and the polarisation large enough to not completely suppress SCAR and the coupling to the leads should be of the order of the superconductor gap $\Gamma, \; \Gamma_S, \; \Gamma_{SF} \approx \Delta$. Nowadays, this coupling is generically obtained in experiments using InAs nanowires or carbon nanotubes as quantum dots \cite{schoene1}.

\section{Conclusion}
To conclude, we have first considered superconductor ferromagnet beamsplitters without a specific consideration of interface properties. We found that splitting of spin-polarized p-wave Cooper pairs can easily be identified in the conductance. However, p-wave superconductors that are usable in experiments are not available but we proposed a scheme to mimic its behavior by taking into account the spin-activity of superconductor-ferromagnet interfaces. The newly identified SCAR process allows to obtain split p-wave Cooper pairs which gives access to the Bell states other than $|\Psi^-\rangle$.\\

The authors would like to thank A. Levy Yeyati, L. Hofstetter and S. Maier for many interesting discussions. The financial support was provided by the DFG under grant No. KO--2235/3, and `Enable fund' of the University of Heidelberg.

\end{document}